# THE APPLICATION OF FUZZY LOGIC TO THE CONSTRUCTION OF THE RANKING FUNCTION OF INFORMATION RETRIEVAL SYSTEMS

## N.O. RUBENS

*University of Massachusetts, Department of Computer Science*
*Amherst, MA 01003, USA*
*1 (413) 545-2744, norouben@nsm.umass.edu*

The quality of the ranking function is an important factor that determines the quality of the Information Retrieval system. Each document is assigned a score by the ranking function; the score indicates the likelihood of relevance of the document given a query. In the vector space model, the ranking function is defined by a mathematic expression such as:

$$score(q,d) = \sum_{t \in q} \text{tf(t in d)} * \text{idf(t)} * \text{getBoost(t.field in d)} * \text{lengthNorm(t.field in d)} * \text{overlap(q,d)} * \text{queryNorm(q)}$$

We propose a fuzzy logic (FL) approach to defining the ranking function. FL provides a convenient way of converting knowledge expressed in a natural language into fuzzy logic rules. The resulting ranking function could be easily viewed, extended, and verified:
- if (tf is high) and (idf is high) → (relevance is high);
- if (overlap is high) → (relevance is high).

By using above FL rules, we are able to achieve performance approximately equal to the state of the art search engine Apache Lucene (ΔP10 +0.92%; ΔMAP -0.1%). The fuzzy logic approach allows combining the logic-based model with the vector model. The resulting model possesses simplicity and formalism of the logic based model, and the flexibility and performance of the vector model.

## 1. Introduction

Information Retrieval (IR) model could be defined formally as a quadruple [$D$, $Q$, $F$, $R(q_i, d_j)$] where $D$ is a set of logical views of documents, $Q$ is a set of user queries, $F$ is a framework for modelling documents and queries, and $R(q_i, d_j)$ is a ranking function which associates a numeric value to the document $d_j$ according to a system assigned likelihood of relevance to a given user query $q_i$ [1]. The quality of the ranking function is an important factor that determines the quality of the IR system.

Logic based Boolean model is one of the earliest models used in the IR systems. The Boolean model owes its former popularity to its clean formalism and simplicity. However, the Boolean model suffers from major drawbacks: binary decision criterion without any notion of a grading scale; difficulty of translating the query into Boolean expressions [1, 6, 4].

Vector model is a popular retrieval model. The main advantages of the vector model are: its term-weighting scheme improves retrieval performance; its partial matching strategy allows retrieval of documents that approximate the query conditions; it's weighting scheme sorts the documents according to their degree of similarity to the query [1]. However, vector model does not possess clean formalism and simplicity.

Fuzzy logic [13] allows combining the logic based model with the vector model. The resulting model possesses simplicity and formalism of the logic based model, and the flexibility and performance of the vector model.

## 2. Related work

Fuzzy logic has not been applied to defining ranking function directly; however, fuzzy set model has been used to define fuzzy queries [2], fuzzy relationships between query terms and documents [7,8]. Each query term defines a fuzzy set and each document has a degree of membership in the corresponding set. The Fuzzy Set model performs query expansion based on principles of fuzzy logic. A thesaurus is constructed by defining a term-term correlation matrix. The correlation matrix is used to define a fuzzy set associated to each index term $k_i$. Document $d_j$ has a degree of membership $\mu_{i,j} = 1 - \prod_{k_i \in d_j}(1 - c_{i,j})$. The procedure to compute the document's relevance given a query is analogous to the procedure used by the Boolean model, except rules of fuzzy logic are used [8]. The fuzzy set model approach is not popular among the information retrieval community and has been discussed mainly in the literature dedicated to fuzzy theory [1]. Recent attempts utilizing fuzzy search were tried at TREC 2001 with the search engine NexTrieve. NexTrieve used a combination of the exact search and fuzzy search. The conference paper that describes NexTrieve [3] unfortunately does not provide details on the theoretical foundation and implementation of the system. It appears that application of the fuzzy logic was to the position, and to the scoring: terms in different parts of the document would get different scores and not all of the words would need to be present in order for the document to get a high score. According to the authors of NexTrieve, one of the biggest drawbacks of the system was that it did not take into account word frequency within a document and document length which has been shown to be a crucial part of the ranking score. Performance of NexTrieve system was substandard with average precision of 0.13; and after some additional modifications (adding word frequency, and document length parameters) were made, it went up to 0.19; which was still substandard [3].

## 3. Motivation

### 3.1. THEORETICAL FIT BETWEEN FUZZY LOGIC MODEL AND INFORMATION RETRIEVAL MODEL

The Information Retrieval system retrieves documents based on a given query. Both the documents and in most cases, the queries, are instances of natural language. Natural langue is often vague and uncertain [9]. It is difficult to judge something that is vague and uncertain with deterministic crisp formulas and/or crisp logical rules. Fuzzy logic is based on the theory of fuzzy sets, a theory which relates to classes of objects with un-sharp boundaries in which membership is a matter of degree [12]. Documents, queries and their characteristics could easily be viewed as fuzzy granular classes of objects with un-sharp boundaries and fuzzy memberships in many concept areas [14].

Fuzzy logic is a logical system, which is an extension of multi-valued logic [12]. Use of fuzzy logic provides the benefits of the Boolean method while overcoming its drawbacks. Since the concept of fuzzy logic is quite intuitive, the fuzzy logic model provides a framework that is easy to understand for a common user of IR system. Documents retrieved by a query are evaluated by the rules of the Fuzzy Inference System (FIS) that have precise semantics. Unlike the Boolean model that is based on binary decision criterion {relevant, not relevant}, fuzzy logic expresses relevance as degrees of memberships (e.g., document | query could have a relevance measure with the following degrees of membership: 0.7 highly relevant and 0.5 moderately relevant and 0.1 not relevant). Fuzzy logic is tolerant of imprecise data [1].

### 3.2. EXPERT KNOWLEDGE TRANSFER PROCESS

Since IR deals with natural language, many of the rules that are used to determine relevance of documents come from experts and from experience. Before the rules are converted to formulas, they are often communicated as observations in natural language (e.g., if most of the terms of the query are present in the document, then the document is likely to be relevant; if a term of a query occurs in a document often, that will increase the likelihood of the document being relevant, etc.). Fuzzy logic allows incorporating rules into the system in a natural way. The basic concept underlying FL is that of a linguistic variable, a variable whose values are words rather than numbers. FL may be viewed as a methodology for computing with words rather than numbers. Even though words are inherently less precise than numbers their use is closer to human intuition. Computing with words allows the tolerance for imprecision [12].

### 3.3. OPTIMIZATION/VERIFICATION PROCESS

Various graphical user interfaces such as Matlab Fuzzy Logic Toolbox provide a convenient way to view all the components of an FIS; to modify them easily; and to examine and verify the effects of changes. Parameters of FIS could be modified systematically by utilizing various optimization approaches. The fuzzy logic approach is a very flexible one. It is possible to make small improvements without disturbing the integrity of the system, by changing parameters of the parts of the system such as rules and membership functions. If more granularity is desired more rules and/or membership functions could be added.

3.4. IMPLEMENTATION

*Baseline Model.* The Ranking Fuzzy Inference System (R-FIS) could be based on various retrieval models that have well defined rules and provide access to underlying features. The vector model was chosen as the baseline model due to its good performance. The R-FIS input variables are typical variables that are used in tf.idf based systems: tf.idf, overlap. It is beneficial to use variables that have been established to be significant in determination of document's relevance, without them, the results would suffer [3].

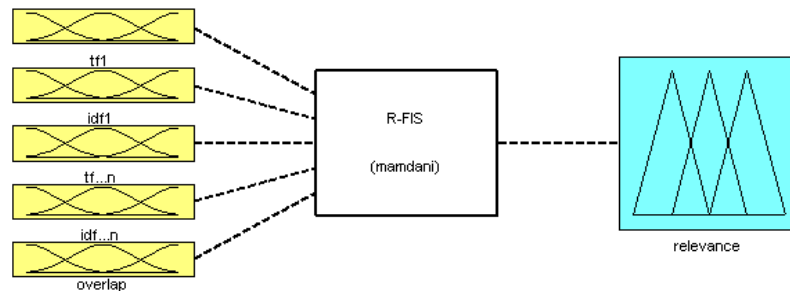

*Figure 1.* Fuzzy Inference System

3.5. FUZZY INFERENCE SYSTEM

*Rules.* Ranking Fuzzy Inference System (R-FIS) is constructed with the use of Matlab Fuzzy Logic Toolbox. The first step in construction of an FIS is to define rules. Rules will be derived from a common knowledge about information retrieval and the tf.idf weighting scheme. The order of the rules in FL does not affect the output.

If many of the terms of the query are found in the document (overlap), the document is likely to be highly relevant:
• if (overlap is high) $\rightarrow$ (relevance is high)
For each of the query terms the following rules are defined:
If a query term in a document has high tf and idf measures, the document is likely to be highly relevant:
• if (tf is high) and (idf is high) $\rightarrow$ (relevance is high)

We have found that the performance of the system is better if the rules that penalize low features are added. To achieve this we added the negated rules for each of the rules above:
• if (overlap is not high) $\rightarrow$ (relevance is not high)
• if (tf is not high) and (idf is not high) $\rightarrow$ (relevance is not high)

Approach to simply negate the rules is compact but it assumes that the opposing membership function is inversely symmetrical. Another approach to creating negated rules is by adding appropriate membership functions such as *low* and *high*.

Each rule has a weight associated with it. In case of R-FIS each tf.idf rule was assigned a weight of 1/t, where t is a number of terms in a query. Weight for the overlap rule was 1/6 of the weight of the tf.idf rule. Overlap rule weight is lower due to the fact that to some degree overlap rule is already represented in each of the tf.idf rules.

*Fuzzy Sets / Membership Functions.* It is necessary to give mathematical meaning to the linguistic variables mentioned in above rules: high relevance, low relevance, high tf, low tf, high idf, low idf. It is necessary to define fuzzy sets (see Figure 2). All input variables and output variables currently have two fuzzy sets associated with each variable: high, not high. If greater granularity is desired, more fuzzy sets could be defined such as for example: very low, low, medium, high, very high, etc. A membership function is a curve that defines how each point in the input space is mapped to a degree of membership of fuzzy set. There are various membership function types such as: sigmoid, Gaussian, trapezoidal (trapmf), triangular (trimf), etc. [5]. Gaussian and sigmoid based functions through a higher number of parameters provide for a higher flexibility.

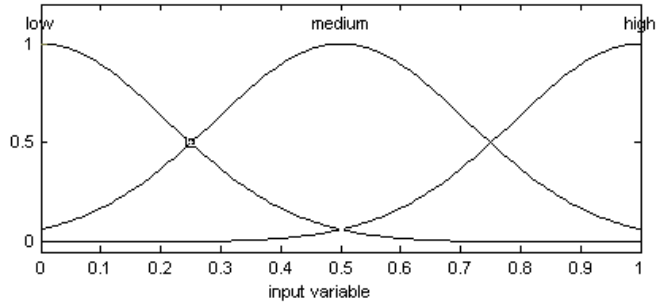

*Figure 2.* Fuzzy Sets / Membership Functions

Generally Gaussian function better models the underlying variables, but knowledge about variables that are being modelled is necessary in order to define parameters of the function appropriately. In this case, linear functions such as trapmf and trimf achieve suitable performance without the need for further tuning. In certain cases, it seems more beneficial to use linear membership functions such as trapezoidal and triangular membership functions. For example, in the case of idf; it appears to be more efficient to use trapmf. Trimmf since idf has already been normalized with log function: $idf = \log(\frac{N}{n})$. For each of the input and output variables of the R-FIS two triangular membership functions were defined: high, not high.

*Fuzzy Inference Process.* The Fuzzy Inference Process is performed automatically, but in order to explain how the system functions, each of the steps will be examined (see Figure 3).

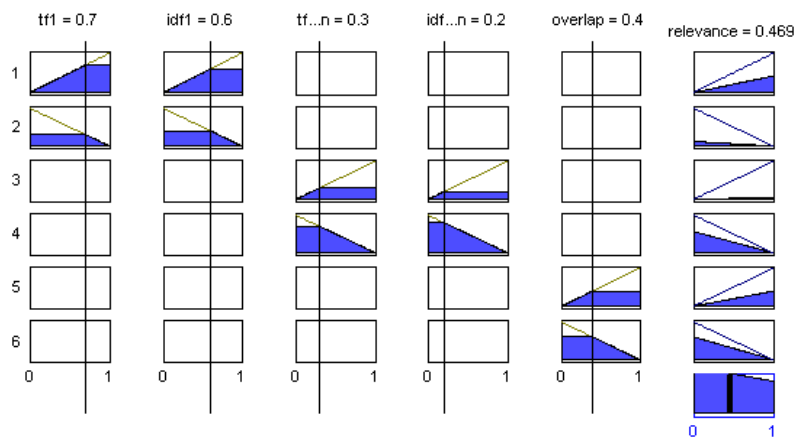

*Figure 3.* Fuzzy Inference Process

*Input Fuzzification.* The first step is to take the crisp numerical values of the inputs and determine the degree to which they belong to each of the appropriate fuzzy sets via membership functions [5]. For example: tf1 = 0.7, this would be translated into 0.7 degree of membership in fuzzy set "high" and 0.3 degree of membership in fuzzy set "not high". Same procedure would be applied to all of the inputs (see Figure 4).

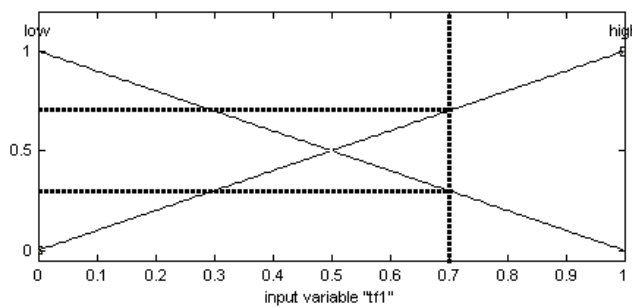

*Figure 4.* Input Fuzzification

23

*Fuzzy Operator Application.* Once the inputs have been fuzzified, the degree to which each part of the antecedent has been satisfied for each rule is known. If the antecedent of a given rule has more than one part, the fuzzy operator is applied to obtain a number that represents the result of the antecedent for that rule [5]. Let's examine rule: if (tf1 is high) and (idf1 is high) → (relevance is high). In this case the input to the fuzzy operator is two membership values from fuzzified input variables: e.g. tf1 has a 0.7 degree of membership in fuzzy set high and idf1 has a 0.6 degree of membership in fuzzy set high. For the R-FIS product (prod) was selected as a fuzzy operator *for* and method so the result is 0.7 * 0.6 = 0.42. This procedure is applied to every rule. *The and fuzzy* operator could be seen as an aggregation applied locally in this case to the terms of the rule. There are different operators *for* and operator such as: min, prod. Prod has a much better theoretical fit, since by using the prod operator, output is determined by all features of the terms and not just the minimum one. *Prod* has also been established to be an appropriate and method realized by a formula based tf.idf model; since in the formula tf and idf are combined through use of product operator:

$$score(q,d) = \sum_{t \in q} tf(t \text{ in } d) * idf(t) * \ldots$$

*Implication Method Application.* The input for the implication process is a single number given by the antecedent, and the output is a fuzzy set. A consequent of the implication method is a fuzzy set represented by a membership function, which weights appropriately the linguistic characteristics that are attributed to it (see Figure 5). The consequent is reshaped using a function associated with the antecedent (a single number) and the weight of the rule. Implication is applied to every rule [5].

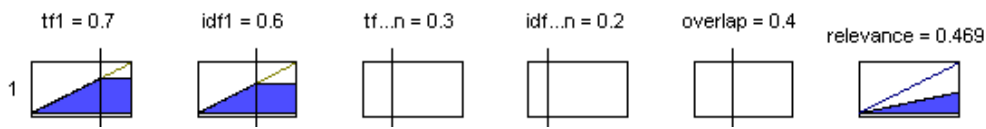

Figure 5. Implication Method

The implication operator determines the shape of the consequent fuzzy set. The prod operator appears to be a better fit in comparison with the min operator; since prod scales the consequent fuzzy set unlike min that truncates the consequent fuzzy set. The prod operator allows the output fuzzy set to retain its shape properties; unlike the min operator that alters the shape of the resulting fuzzy set.

*Output Aggregation.* Since decision is based on all of the rules in the FIS, the rules must be combined in order to make the decision. Aggregation is the process by which the fuzzy sets that represent the outputs of each rule are combined into a single fuzzy set. The input of the aggregation process is the list of fuzzy sets that represent the outputs of each rule (see Figure 6). The output of the aggregation process is a fuzzy set [5].

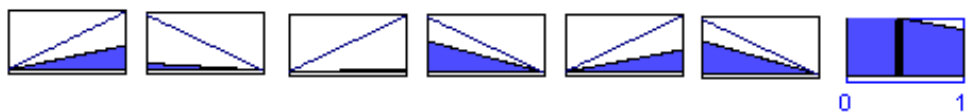

Figure 6. Output Aggregation

There are a number of different aggregation methods available, such as: max, sum, probabilistic or. The nature of the information retrieval dictates that the determination of the ranking should be done based on all of the rules. In this case the sum aggregation method appears to be a much better fit. Sum has also been established to be an appropriate aggregation method by a formula based tf.idf model since terms are combined through the sum operator: $score(q,d) = \sum_{t \in q} \ldots$ .

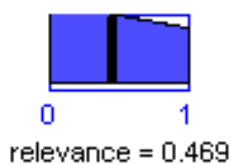

Figure 7. Output Defuzzification

*Output Defuzzification.* The input for the defuzzification process is the aggregate output fuzzy set and the output is a single number (see Figure 7). Fuzziness helps the rule evaluation during the intermediate steps; however the final desired output for each variable is generally a single number. Fuzzy set must be defuzzified in order to resolve a single output value from the set. There are various methods for defuzzification such as: centroid, bisector, middle of maximum (the average of the maximum value of the output set), largest of maximum, and smallest of maximum.

Centroid method is the most widely used method. In this case the centroid method is used, since it satisfies the underlying properties of the system and exhibits the best performance. The centroid method returns the centre of the area under the curve. In this case it is 0.469.

**4. Evaluation**

4.1. EVALUATION DATA SET

To evaluate the effectiveness of R-FIS, data from the NIST TREC 2004 Robust Retrieval Track was used. The robust retrieval track explores methods for improving the consistency of retrieval technology by focusing on poorly performing topics [10].

TABLE 1. Document Corpus

| Source | # Docs | Size (MB) |
|---|---|---|
| Financial Times | 210,158 | 564 |
| Federal Register 94 | 55,630 | 395 |
| FBIS, disk 5 | 130,471 | 470 |
| LA Times | 131,896 | 475 |
| *Total Collection:* | *528,155* | *1904* |

The Robust test set contains 250 topics: topics 301-450 (ad hoc topics from TREC 6-8), topics 601-650 (new topics for 2003 robust track), and topics 651-700 (new topics for 2004 robust track) [11].

TABLE 2. Relevant document statistics for topic sets

| Topic Set | Number of topics | Mean Relevant per topic | Minimum Number Relevant | Maximum Number Relevant |
|---|---|---|---|---|
| Old | 200 | 76.8 | 3 | 448 |
| New | 49 | 42.1 | 3 | 161 |
| Hard | 50 | 88.3 | 5 | 361 |
| Combined | 249 | 69.9 | 3 | 448 |

*Baseline System.* Apache Lucene version 1.4.3 [15] with query expansion module [16] was used as the baseline search engine. Lucene is an open source, high-performance, full-featured text search engine written entirely in Java.
*System Configuration.* FIS Rules:
• if (overlap is high) → (relevance is high)
For each of the query terms:
• if (tf is high) and (idf is high) → (relevance is high)
• if (tf is not high) and (idf is not high) → (relevance is not high)

TABLE 3. FIS Specifications

| Fuzzy Inference System Type | Mamdani |
|---|---|
| And method | product |
| Implication method | product |
| Aggregation method | sum |
| Defuzzification | centroid |
| Membership Function Type | triangular [0,1] |

Each rule has a weight associated with it. Each query term rule assigned a weight of 1/t, where t is a number of terms in a query. Weight for the overlap rule is 1/6 of the weight of the query term rule, due to the fact that to some degree overlap rule is already represented in each of the query term rules (see Figure 8).

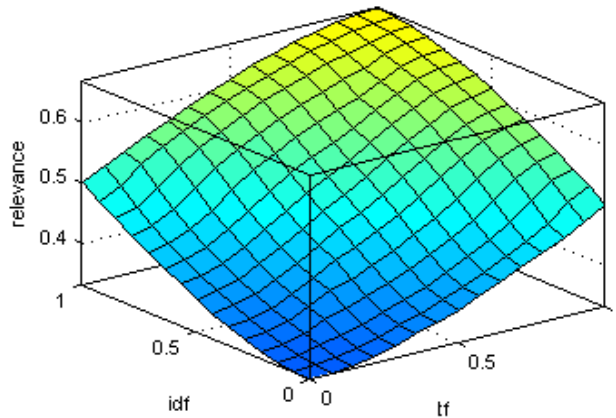

*Figure 8.* Rule Surface

## 4.2. PERFORMANCE EVALUATION

TABLE 4. Evaluation Results

| Tag | Topic Set | MAP | P10 | %no |
|---|---|---|---|---|
| Lucene | Old | 0.2232 | 0.3945 | 15.00% |
| R-FIS | Old | -0.0061 | +0.0035 | -0.00% |
| Lucene | New | 0.2738 | 0.4143 | 10.20% |
| R-FIS | New | +0.0201 | +0.0326 | -2.00% |
| Lucene | Hard | 0.1374 | 0.286 | 28.00% |
| R-FIS | Hard | +0.0117 | +0.026 | -2.00% |
| Lucene | Combined | 0.2332 | 0.3984 | 14.10% |
| R-FIS | Combined | -0.001 | +0.0092 | -0.40% |

*Values given are the mean average precision (MAP), precision at rank 10
averaged over topics (P10), the percentage of topics with no relevant
in the top ten retrieved (%no).*

R-FIS has slightly outperformed Lucene on all of the topic sets on the P10 and %no measures. MAP measure was slightly better for R-FIS on *New* and *Hard* topic sets and slightly worse for *Old* and *Combined* topic sets. Overall, we believe R-FIS performed very well, considering that Lucene is a state of the art vector based search engine.

**Conclusion**

We presented a new method of defining the ranking function by combining the logic based model with the vector model through the use of fuzzy logic. Fuzzy logic provides a convenient way of converting existing knowledge into fuzzy logic rules. To construct a system with a good performance, a basic understanding of the tf.idf principles and the basic principles of information retrieval theory appears to be sufficient. The resulting model possesses simplicity and formalism of the logic based model, and the flexibility and performance of the vector model.

For the baseline model we have used vector based model due to its speed and performance, but fuzzy logic approach could be applied to any retrieval model that has well defined rules and provides an access to underlying features.

**Acknowledgments**

The author would like to thank James Allan and William Kilmer for helpful discussions and suggestions. I would also like to thank *Open Source* community for providing the numerous tools and systems I have used to produce both my results and this paper.

**Information systems**